\documentclass[conference]{IEEEtran}
\IEEEoverridecommandlockouts
\usepackage{threeparttable}
\usepackage{caption}
\usepackage{pifont}
\usepackage{ntheorem}
\usepackage[nocompress]{cite}
\usepackage{url}
\usepackage{graphicx}
\usepackage{epstopdf}
\usepackage{epsfig,subfigure}
\usepackage{amssymb,amsmath}
\usepackage{enumerate}
\usepackage{multirow}
\usepackage{array}
\usepackage{cite}
\usepackage{mathrsfs}
\usepackage{amsmath}

\usepackage[noend]{algpseudocode}
\usepackage{algorithmicx,algorithm}
\usepackage[justification=centering]{caption}
\usepackage{soul}
\soulregister\cite7 
\soulregister\ref7 
\usepackage{color}
\usepackage{url}

\begin{document}
\title{RingAda: Pipelining Large Model Fine-Tuning on Edge Devices with Scheduled Layer Unfreezing

\author{\IEEEauthorblockN{
Liang Li\IEEEauthorrefmark{1},
Xiaopei Chen\IEEEauthorrefmark{1},
and Wen Wu\IEEEauthorrefmark{1}
}                                     
\IEEEauthorblockA{\IEEEauthorrefmark{1}Frontier Research Center, Peng Cheng Laboratory, Shenzhen, China}

Email:{\{lil03, chenxp, wuw02\}@pcl.ac.cn}\\
}
\thanks{\scriptsize{This work was supported in part by NSFC 62201071, in part by the Peng Cheng Laboratory Major Key Project under Grants PCL2023AS1-5 and PCL2021A09-B2, in part by NSFC 62201311, and in part by the Young Elite Scientists Sponsorship Program by CAST under Grant 2023QNRC001.}}
}

\maketitle

\begin{abstract}
To enable large model (LM) based edge intelligent service provisioning, on-device fine-tuning with locally personalized data allows for continuous and privacy-preserving LM customization. In this paper, we propose RingAda, a collaborative training framework designed for fine-tuning transformer-based LMs on edge devices. Particularly, RingAda performs parameter-efficient adapter fine-tuning across a set of interconnected edge devices, forming a ring topology for per-batch training by sequentially placing frozen transformer blocks and their trainable adapter modules on the devices. RingAda follows a novel pipeline-parallel training mechanism with top-down adapter unfreezing, allowing for early-stopping of backpropagation at the lowest unfrozen adapter layer, thereby accelerating the fine-tuning process. Extensive experimental results demonstrate that RingAda significantly reduces fine-tuning time and memory costs while maintaining competitive model performance compared to its peer designs.


\end{abstract}

\IEEEpeerreviewmaketitle

\section{Introduction}
Transformer-based large models (LMs), like LLaMa \cite{touvron2023llama}, and GPT~\cite{brown2020language}, have brought about significant progress in artificial intelligence. Their ability to understand context and nuances allows them to handle various tasks in natural language processing (NLP), computer vision (CV), and many other applications more effectively. LMs trained on large datasets can generalize and apply their knowledge to new tasks by fine-tuning domain-specific samples, such as user reviews and messages. To better use the geo-distributed data with privacy preservation, there is an increasing demand for collaboratively fine-tuning LMs on edge devices with their local data~\cite{Li2021infocom}. This is supported by the continuous evolution of modern edge devices with on-device computing capabilities tailored for mobile AI tasks, bringing us closer to more intelligent and personalized assistive agents on edge devices~\cite{li2023energy}.

Towards a privacy-friendly LM training paradigm, recent efforts have primarily focused on the federated learning (FL) paradigm for distributed fine-tuning~\cite{chen2022fedtune}, referred to as FedLLM. Targeting a specific downstream task, FedLLM follows the typical FL protocol where participating edge devices load the pre-trained LM and perform local updates with periodical aggregation. However, despite the potential system- and hardware-support, fine-tuning LMs on edge devices under the FedLLM workflow faces significant barriers in memory and computational capacity~\cite{shi2022toward}. The prevalent LM training algorithm requires substantial memory for storing intermediate results like activations and optimizer states~\cite{xu2023FwdLLM}. The limited computing capacities of edge devices directly affect real-time response and fine-tuning speed. For example, BERT-large has 330M trainable weights and takes 250,000 PFLOPs to train, which is 6×/23× higher than ResNet-152, respectively~\cite{devlin2018bert}. Fine-tuning BLOOM-176B with Adam would require almost 3 TB of GPU memory to store the model, gradients, and optimizer states, while the typical edge devices possess only 4–12GB DRAM. For LMs, fine-tuning all model parameters on the downstream task, as a traditional way in deep neural networks, becomes inefficient, or even impractical.

For efficient LM fine-tuning, adapters~\cite{houlsby2019adapter} - a recently proposed parameter-efficient fine-tuning (PEFT) technique for both CV and NLP tasks - can significantly reduce the trainable parameters while achieving competitive performance compared to full fine-tuning. Its key idea is to freeze the whole original model but insert a few small modules into different locations inside it (e.g., between Transformer layers). There are some recent attempts incorporating PEFT techniques into FedLLM (e.g., AdaFL~\cite{cai2023adaFL}, FedPrompt~\cite{zhao2023fedprompt}) to accelerate the training and see improvements in saving the network traffic between devices and aggregator due to the remarkably reduced size of model updates. Yet they do not fully address the issues of excessive memory footprint. Despite their advantages in parameter efficiency, most PEFT methods necessitate the storage of intermediate activations, similar to the requirements of full fine-tuning, to calculate the gradients of the trainable parameters. Typically, they consume more than 70\% activation memory of full fine-tuning~\cite{liao2024make}. Since activations significantly contribute to the memory requirements during training, on-device fine-tuning a pre-trained LM with PEFT is still not feasible due to memory constraints.  

To address these challenges, we propose RingAda, an edge collaborative training framework designed for fine-tuning transformer-based LMs with pluggable adapters on edge devices. RingAda utilizes layer-wise model partitioning to sequentially place the LM's transformer blocks and their associated adapter modules on multiple edge devices while retaining copies of the embedding layer (for input) and the head layer (for output) locally on each device. It then forms a ring topology with dynamic start and end points for per-batch training, enabling a round-robin type training workflow that leverages local data from all participating devices. Particularly, RingAda allows the deactivation of the bottom-layer adapters to early-stop backward propagation, based on which it pipelines and paralyzes the training iterations without model staleness, thereby accelerating the fine-tuning. We evaluate the performance of RingAda with extensive experiments, and the experimental results demonstrate that our design can remarkably reduce the latency for LM fine-tuning on edge devices at a limited cost.

Compared with prior work on distributed LM edge training, to the best of our knowledge, RingAda is the first to enable adapter-based pipeline parallel fine-tuning on multiple edge devices, optimizing memory and flops usage while leveraging all local data contributions to LM fine-tuning without label sharing. The remainder of the paper is organized as follows. Section~II gives a brief introduction to adapter fine-tuning and pipeline parallel training. Section~III overviews our RingAda design. Section~IV elaborates on the pipeline parallel training process. Section~V presents the performance evaluation, and Section~VI concludes the paper.

\section{Preliminaries}

\subsection{Adapter-based LM Fine-tuning}

Standard full fine-tuning incurs significant computational costs and may potentially compromise the model's generalization ability. As one of the emerging PEFT techniques, adapter~\cite{ houlsby2019adapter } maintains the pre-trained backbone parameters frozen and inserts tiny tunable adapter modules (nearly 2\% of total number parameters) within Transformer blocks. During fine-tuning for a specific downstream task, only the weights of these adapter modules are updated. Adapter fine-tuning is often competitive with full model fine-tuning and may help stabilize the convergence process to avoid overfitting~\cite{ ruckle2021adapterdrop }. 

\begin{figure}\centering 
  \centering
  \includegraphics[width=.4\textwidth]{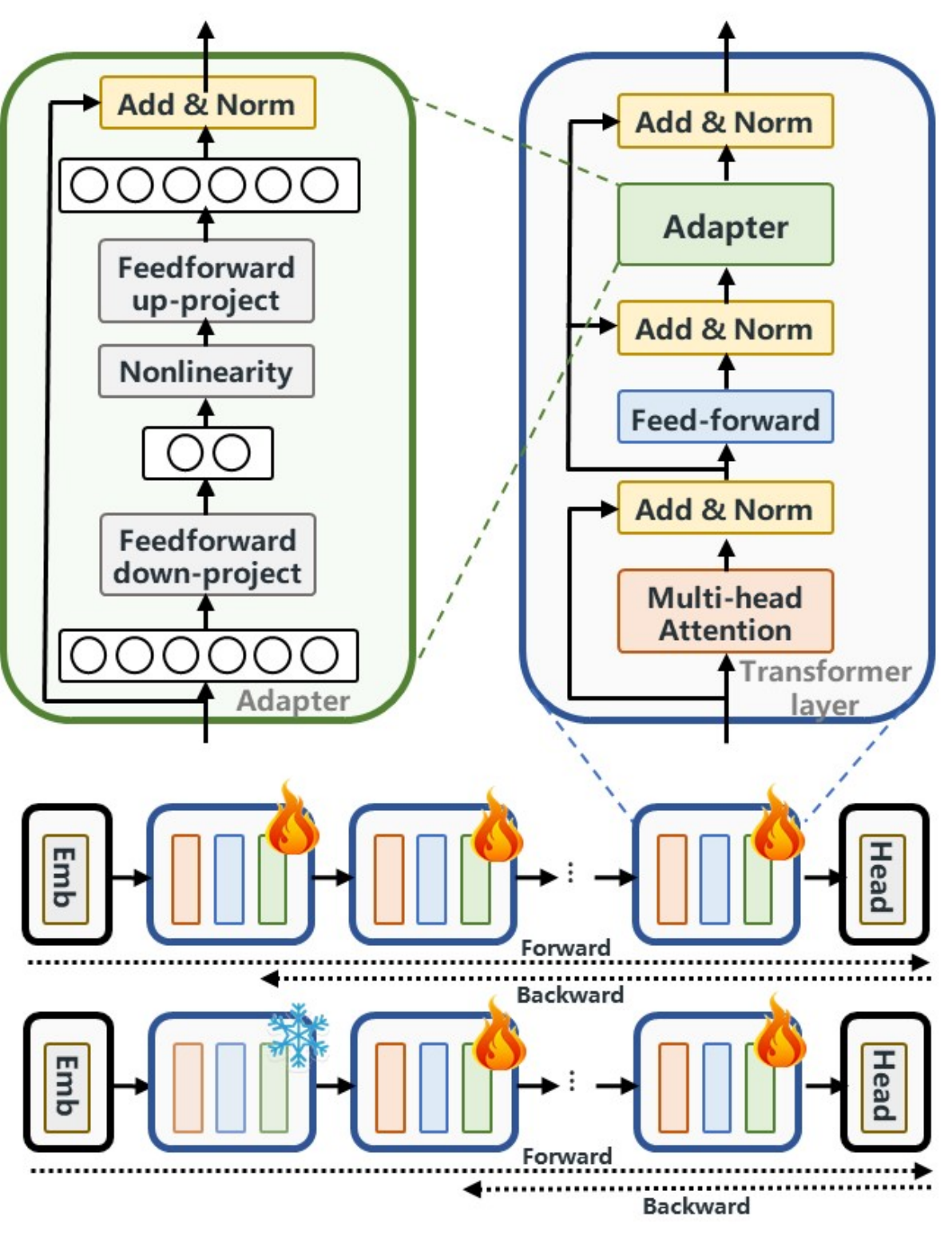}\\
  \caption{Serial adapter architecture with layer freezing for transformer-based large models.}\label{fig:adapter}
\end{figure}

Fig.~\ref{fig:adapter} illustrates the serial adapter method, following a state-of-the-art adapter variant to insert only one adapter module after the second sublayer, i.e., the feed-forward network ``add \& layer norm'' sublayer. Typically, an adapter module consists of a down-projection matrix $\textbf{W}_{down}\in \mathbb{R}^{n*m}$ to project the input to a lower-dimensional space, followed by a non-linear activation function $\sigma(\cdot)$, and an up-projection matrix $\textbf{W}_{up}\in \mathbb{R}^{m*n}$. Here, $n$ represents the dimension of the hidden layer, and $m$ serves as the bottleneck dimension, a hyperparameter used in configuring the adapters. Denoting $\textbf{h}_{in}$ as the input to the adapter, the computation within the adapter module (with residual) can be summarized as follows:

\begin{equation}
\textbf{h}_{in} \leftarrow \textbf{h}_{in}+\sigma(\textbf{h}_{in} \textbf{W}_{down}) \textbf{W}_{up}.   
\end{equation}

Recent studies reveal that instead of injecting adapters into all the transformer layers, freezing or removing adapters from lower transformer layers (i.e., AdapterDrop) can reduce the computational overhead with a neglectable decrease in fine-tuning performances~\cite{liu2023improving}. The reason behind is that the bottom layers of the pre-trained foundation model usually learn low-level coarse feature representations that are mostly shared with the downstream tasks, and thus require minor adjustments during fine-tuning. Strategically removing lower-layer adapters allows backpropagating through as few layers as possible, which may open up opportunities to further improve the efficiency of training adapters.

\begin{figure*}\centering 
  \centering
  \includegraphics*[width=.95\textwidth]{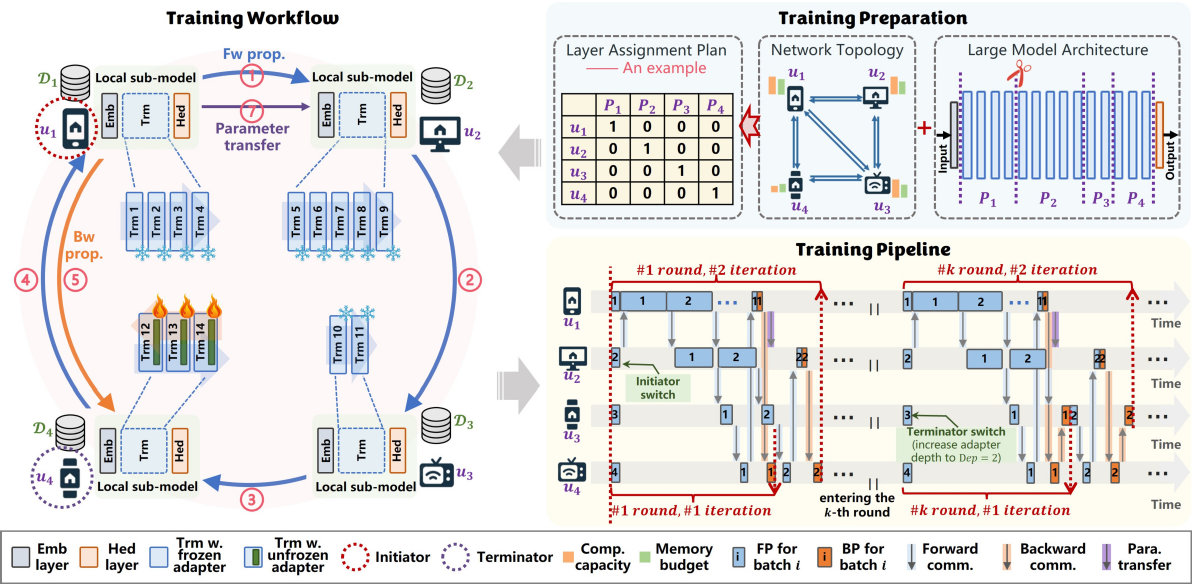}\\
  \caption{The training workflow of RingAda (an instance with four edge clients).}\label{fig.workflow}
\end{figure*}

\subsection{Pipeline Parallel Training}

Pipeline parallelism reduces the computational burden and memory footprint of individual edge devices by dividing a pre-trained model into several sub-models and assigning each device with one. This allows even devices with very low memory budgets to participate in the training process. Some recent works extend the pipeline parallel training to edge computing scenarios like smart homes, where a set of edge devices within a local area network collaboratively train a model for specific tasks. A pioneering effort is Confidant~\cite{chen2024confidant}, a multi-backend edge collaborative training framework, which dynamically partitions the model into several sub-models and trains each of them on a different edge device in a pipeline parallel manner, fitting the devices’ real-time available computational capacity and peer-to-peer communication links. In the \textit{Confidant} system, the data samples required for fine-tuning the pre-trained LM are stored at one specific device, which serves as the head of the training pipeline to hold several lower layers of the model. Notice that pipelining schemes need to ensure that inputs see consistent weight versions across forward and backward passes for well-defined synchronous weight update semantics. For a certain batch, the forwarding and the backwarding computations in \textit{Confidant} are executed by traversing the sub-models on different devices in consecutive order, applying the one-forward-one-backward (1F1B) rule to avoid disorder in model updates. During the forward pass, each device applies its subset of layers to the inputs supplied by the previous stage, and then sends the outputs of the last layer to the next stage. For the backward pass, this process is reversed, with each pipeline stage passing the gradients to the same device that previously supplied it with input activations. 

Particularly, existing pipeline parallelism based collaborative training mechanisms allow batch-level parallel computation across devices to accelerate the training, where multiple batches of data samples are allowed to continuously feed in the model and each device runs in parallel on a different batch of inputs. In this way, each device is allowed to forward a batch using stale sub-model weights instead of waiting for the newest ones. This necessitates storing the copy of the weights utilized to forward a certain batch for subsequent backwarding of the same batch. It could incur significant memory storage costs, rendering it may not work well or work at all in edge LM fine-tuning scenarios where mobile devices already face memory pressure when handling LM-related learning tasks.

\section{RingAda Design}

\subsection{Overview}

The RingAda system consists of a set of distributed mobile clients (e.g., IoT devices in a smart home) denoted by $\mathcal{U}=\{1,2,...,u,...,U\}$, which collaboratively fine-tune a transformer-based LM for a specific downstream intelligent task. Each client $u $ maintains a locally collected dataset $\mathcal{D}_u $ with the size of $|\mathcal{D}_u|$. The goal of collaborative fine-tuning is to learn a unified model that achieves uniformly good performance over all the clients. 

A control node (i.e., coordinator) is needed to collect global information about system/learning statuses while controlling the training process. The clients can communicate with the coordinator to upload system/training status information and receive the scheduling policy during the training. Unlike the parameter server in the FL system, the coordinator does not aggregate the model updates and hence will not become the bandwidth bottleneck. Furthermore, any client who is interconnected with all the clients can act as the coordinator. Since the sizes of system information and control signaling are much smaller than those of model parameters, it is reasonable to ignore the cost (e.g., bandwidth consumption and time cost) for information interaction.

As the backbone of many state-of-the-art LMs (e.g., BERT, GPT-3, ViT, DALL-E 2, Codex, and Gopher), Transformer architecture~\cite{vaswani2017attention} prompts the success of generative artificial intelligence. For simplicity and without loss of generality, in this paper, we discuss the LM based on Transformer as the underlying model. As shown in Fig.~\ref{fig:adapter}, pre-trained foundation models based on Transformer architecture can be divided into the embedding layer (\textit{Emb}) for feature extraction, the multiple Transformer layers (\textit{Trm}) for self-attention-based learning, and the multilayer perceptron (MLP) head layer (\textit{Hed}) for output judgment. Adapter fine-tuning adds learnable adapter modules to the rear of each \textit{Trm}, which are learned together with the \textit{Hed} during fine-tuning~\cite{houlsby2019adapter}. 


RingAda is built upon the serial collaborative computation framework. Before launching the fine-tuning task, the coordinator makes a schedule for partitioning the LM and deploying the sub-models on the clients to build a RingAda system. After the LM partition, the tunable modules of the LM are transferred to the edge devices, where each client hosts a copy of \textit{Emb} and \textit{Hed}. In addition, each client also deploys a continuous portion of middle \textit{Trm} blocks as well as the associated adapter modules, undertaking corresponding forward and backward computing workloads. Thus, the clients form a ring topology, where adjacent clients can communicate with each other through direct communication technologies such as device-to-device (D2D) communication. 

In a training round, the forward and backward propagation of a client is conducted cooperatively by the clients along the ring topology. To save resource consumption at early training stages, we apply the top-down adapter unfreezing policy, where only the parameters of the unfrozen adapters are trained and updated during fine-tuning. The coordinator configures the unfreezing depth, which determines the number of adapter layers that clients need to perform parameter updates during each training iteration. Subsequently, the forward propagation begins with the client who holds the sample batch in the current iteration and traverses the ring topology ended at the client itself to use its locally-stored data label, where each client in the ring topology is responsible for propagating several \textit{Trm} layers. This eliminates the need for devices holding local data to share corresponding labels with other devices, ensuring data privacy and reducing interaction overhead. The backward propagation is performed by each client to compute the gradients of the corresponding layers, where the propagation is early-stopped at the adapter layer specified by the scheduled unfreezing depth. As the example in Fig.~\ref{fig.workflow}, if there are four clients with $4:5:2:3$ \textit{Trm} assignment and the adapter unfreezing depth is set to $3$. The model of $u_1$ is trained by traversing $u_1 \rightarrow u_2 \rightarrow u_3 \rightarrow u_4 \rightarrow u_1$ for forward propagation (blue arrows) and $u_1 \rightarrow u_4 $ for backward propagation (orange arrows). Then, the model of $u_2$ is trained based on the updated sub-models on the clients. Note that the forward propagation before the unfreezing depth layer is independent of the training process of the $u_1$’s model and can be conducted simultaneously to enable training parallelism. 

\begin{algorithm}[!t]
\caption{RingAda Algorithm} \label{AdaConfig}
\hspace*{0.02in} {\bf Initialization:} Unfreezing depth $d$. \\
\hspace*{0.02in} {\bf Input:} Edge device set $\mathcal{U}$, edge devices' state information $(\mathbf{R}_u, C^{comp}_u, C^{mem}_u), \forall u$, layer unfreezing interval $k$. 
\begin{algorithmic}[1]
    \Statex {/* \textbf{On coordinator:} */}
    \State Determine and broadcast the layer assignment plan.
    \State Dispatch the pre-trained layer parameters.
    
    \For{Training round $r = 1, 2, 3, \dots$}
        \Statex {/* \textbf{On edge device:} */}
        \For{Initiator device $u \in \mathcal{U}$}
            \For{Local iteration $i = 1, 2, \dots, I$}
                \State Device $u$ updates the local \textit{Hed} using the latest weights.
                \State Device $u$ samples a mini-batch from its local dataset.
                \State Perform Fw propagation by traversing the ring topology.
                \State $u$ calculates the \textit{loss} and updates the head.
                \State Perform Bw propagation ending at the $(L - d + 1)$-th adapter by reversely traversing the ring topology.
            \EndFor
        \EndFor
        \State Device $u$ sends \textit{loss} to the coordinator.
        
        \Statex {/* \textbf{On coordinator:} */}
        \If{model has converged}
            \State Break
        \ElsIf{$r \mathsf{mod} k = 0$}
            \State $d = d + 1$
            \State Broadcast the layer unfreezing strategy.
        \EndIf
    \EndFor
    \State \Return The fine-tuned LM.
\end{algorithmic}
\end{algorithm}

\subsection{RingAda Tranining Process}\label{sec:TrainProcess}

\begin{table*}[ht!]\centering
\caption{Performance Comparison of Different Schemes}
\begin{threeparttable} 
\renewcommand\arraystretch{1.5}
\begin{tabular}{c|c|c|c|c|c}
\hline
\textbf{Schemes} & \textbf{Memory Usage (MB)} & \textbf{Epochs to Convergence} & \textbf{Convergence Time (s)} & \textbf{Accuracy (F1 Score)} & \textbf{Accuracy (EM Score)} \\ \hline
Single           & 1035.04                    & 600                     & 5103.60                       & 80.0848                           & 70.5881                           \\ \hline
PipeAdapter      & 432.576                    & 640                      & 2428.72                       & 78.6117                           & 68.5741                          \\ \hline
RingAda (ours)   & 373.056                    & 700                     & 1793.18                       & 77.3379                           & 66.8684                           \\ \hline
\end{tabular}
\begin{tablenotes} 
		\item *Memory usage is measured on a per-device basis. 
     \end{tablenotes} 
\end{threeparttable} 
\end{table*}

Without loss of generality, we present the detailed training process of RingAda for an arbitrary number of clients. The \textit{Trm} blocks owned by client $u$ are specified by two functions $\beta(u)=b$ and $\epsilon(u)=e$, indicating that the $b$-th to $e$-th \textit{Trm} blocks are assigned to client $u$. The sub-model for client $u$ in training round $r$ is denoted by $\mathcal{W}^r_{(\beta(u), \epsilon(u))}$. Let $Dep^r$ and $\mathcal{W}^r_{(\beta(u), \epsilon(u))}$ be the adapter unfreezing depth in training round $r$. For clients satisfying $\boldsymbol{1}(\epsilon(u) \leq N_{TM}-Dep^r)>0$, they do not need to perform backward propagation in round $r$; instead, they continuously perform the forward pass for consecutive data batches.  For simplicity, we refer to the client whose local data is sampled for learning in a training iteration as the \textit{"initiator"}, and refer to the client who owns the most bottom unfrozen adapter layer as the \textit{`` terminator’’} (i.e., satisfying $\boldsymbol{1}(\beta(u) \leq N_{TM}-Dep^r+1 \leq\epsilon(u))>0$).

1)	\textbf{Initialization:} In the initialization stage, each client uploads its state information $(\mathbf{R}_u, C^{comp}_u, C^{mem}_u)$ to the coordinator. Here $\mathbf{R}_u=\{R_{u,u’}\}_{\forall u,u’\in \{1,2,…,U\}}$ is a collection of data rate for the link $ u’\rightarrow u$, and $ C^{comp}_u $ and $ C^{mem}_u $ are the computational speed and memory budget of client $u$, respectively. The coordinator determines the layer assignment policy based on the collected system status information and distributes the pre-trained layer parameters to the corresponding clients. The first \textit{initiator} in each training round is selected and informed by the coordinator. The coordinator also sends the necessary training setups to the clients (including the learning rate, number of local iterations, and batch size) to initialize the RingAda training.

2)	\textbf{Forward and Backward Propagation}: In every training round, the selected \textit{initiator} (client $u$) samples a mini-batch data $(\boldsymbol{x}_u,\boldsymbol{y}_u)$ from its local dataset $D_u$. Client $u$ then feeds $\boldsymbol{x}_u $ into its locally loaded \textit{Emb} to get the tokens and positional embeddings, which are sent to the next client possessing the lowest \textit{Trm} block(s). The intermediates from each client’s sub-model are traversed along the ring topology, and the output of the last Transformer block $\overline{\boldsymbol{x}}_u $ is sent back to client $u$ to calculate the training loss, finalizing the forward propagation for the data batch $(\boldsymbol{x}_u,\boldsymbol{y}_u)$. Similar to the forward propagation, the backward propagation is performed over the clients by relaying the gradients in reverse along the ring topology, but early stopping at the \textit{ terminator}. Based on the locally calculated gradients, the \textit{initiator} updates the Head layer while other clients with unfrozen adapters update the parameters of their adapter layers. It should be noted that client $u$ is allowed to continuously sample mini-batch and feed into the distributed deployed LM to initiate the forward propagation of a new training iteration, enabling a pipelined parallel training process over the clients. As such, the 1F1B~\cite{narayanan2019pipedream} training rule is broken for clients whose adapters are all frozen in the current round, while clients with unfrozen adapters strictly follow the classical interleaved forward and backward passes on the same mini-batch. As a result, the forward pass for a given data batch may pause at a client in the ring topology whose unfrozen adapters have not yet been updated through the backpropagation of the previous iteration. 

\begin{figure} \centering
 \subfigure[Training loss vs epochs.\label{fig:conv-curve}]
  {\includegraphics[width=0.829\linewidth]{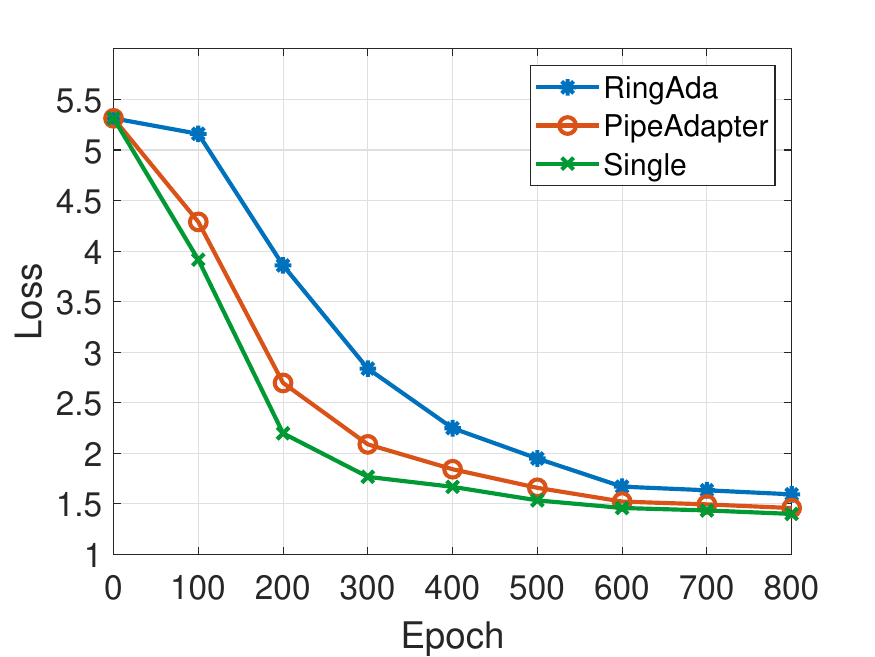}}
  \subfigure[Training accuracy vs time.\label{fig:loss-curve}]
  {\includegraphics[width=0.829\linewidth]{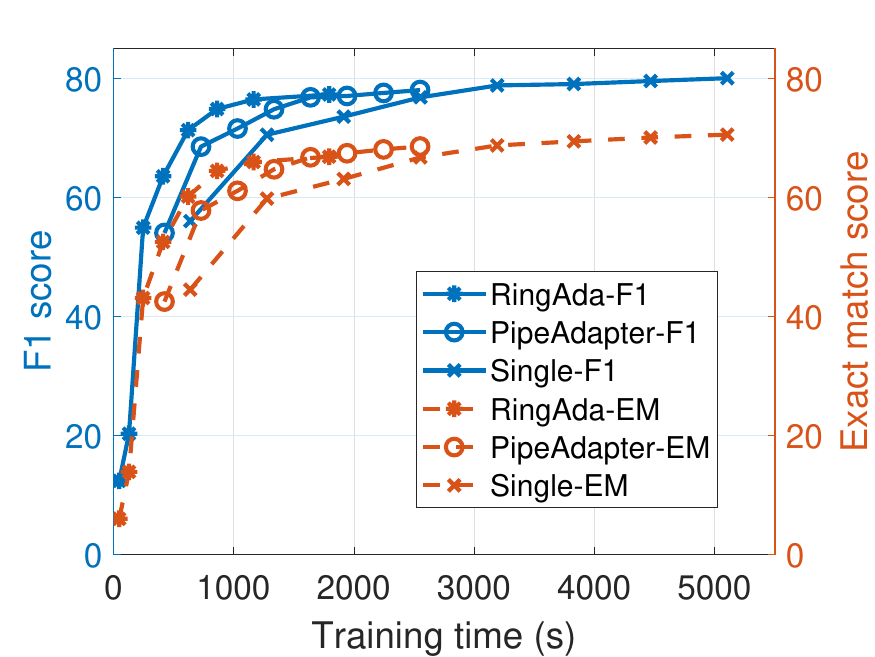}}
  \caption{Training Performance of the three schemes.} 
\end{figure}

3)	\textbf{Complete a training round:} Upon completing a pre-selected number of local iterations, client $u$ designates the next \textit{initiator} $u’$ with the best channel quality with client $u$, and transfers the latest parameters of the \textit{Hed} to $u’$ for updating its local \textit{Hed}. Client $u$ also reports the local loss value to the coordinator to assess training convergence. The training round concludes after all participating clients have acted as the \textit{initiator}. 

The above round-robin cooperative training procedures are repeated across rounds until convergence, as outlined in Algorithm 1. RingAda offers three-fold advantages over existing cooperative fine-tuning frameworks. First, RingAda integrates adapter fine-tuning into the pipeline parallel training framework, rather than full model fine-tuning, which helps alleviate computation burdens on mobile devices and accelerate convergence. Second, RingAda enables the utilization of decentralized data from multiple mobile devices for jointly fine-tuning a uniform LM without the need for label sharing, improving both data efficiency and privacy. Finally, RingAda supports parallel training through scheduled layer freezing, avoiding the propagation of multiple data batches on stale model weights. This ensures all parameters are updated with the latest model, resolves the model staleness issue, and reduces memory pressure by eliminating the need to store multiple versions of the model.

\section{Performance Evaluation}

In our evaluations, we leverage mBERT, an established pre-trained multilingual model, for question-answering tasks using the SQuAD dataset \cite{rajpurkar2016squad}. We use language adapters available in the AdapterHub, which follows the adapter configurations from MAD-X for cross-lingual transfer. The reported results are averaged across 4 runs on RTX3090 GPUs. We use the PyTorch library to implement the RingAda fine-tuning and use trace-based simulation to evaluate the performance. We profile the computation time of forward and backward propagation on different edge devices by scaling the computational speed, which can be easily measured using the \textit{torch.nn} package. The forward and backward propagation time of different layers on different computing capacities is recorded in a lookup table.

RingAda performs top-down unfreezing during fine-tuning. We fine-tune with the head layer and the top-most adapter unfrozen, and for every 40 steps, we unfreeze the next adapter and continue. We set the RingAda system to consist of $4$ edge devices. We compare RingAda with the following baselines: 1) \textit{PipeAdapter} keeps all the adapters unfrozen for fine-tuning throughout the fine-tuning process, and the devices follow pipeline training procedures. 2) \textit{Single} applies the classic adapter fine-tuning scheme relying on a single device, with all the adapters unfrozen.

Fig. 3(a) compares the convergence curves of the three fine-tuning schemes over 800 epochs. Due to the partial unfreezing of trainable adapters, RingAda exhibits a relatively slower initial convergence rate compared to the baseline, though the gap in losses narrows as training progresses. Fig. 3(b) demonstrates demonstrates that RingAda accelerates convergence, completing 800 epochs in 1793.18 seconds, while \textit{PipeAdapter} and \textit{Single} take 2428.72 seconds and 5103.60 seconds, respectively. This speedup is achieved despite the additional inter-device communication time in RingAda compared to \textit{Single}. The acceleration advantage stems from two key mechanisms: multi-batch parallel training and gradual adapter unfreezing. First, RingAda allows continuous sample mini-batches and feeds them into the LM for forward propagation to enable a pipelined training process across clients, effectively collapsing part of the computation time. Second, backpropagation occurs in reverse along the ring topology, but halts early at the client with the lowest unfrozen adapter layer, thereby skipping the computation and transmission delays associated with backpropagating through the remaining layers. 

We compare the memory usage, the number of epochs to convergence, convergence time, and test accuracy of the three fine-tuning schemes under two transmission rate levels, as shown in Table I. The average memory usage per device is measured using the Python memory profiler. Table I shows that RingAda accelerates the fine-tuning process while maintaining similar final accuracy. RingAda, \textit{PipeAdapter}, and \textit{Single} converge in 1793.18s, 2428.72s, and 5103.60s, respectively. This convergence speed advantage could be more pronounced under favorable communication conditions, where device computation time dominates overall training time. By scheduling layer unfreezing, RingAda effectively modifies the computation graph for forward and backward propagation, optimizing the computational workload on each device. In terms of memory usage, RingAda exhibits much lower memory requirements per device, making it more suitable for memory-constrained edge environments. In contrast, the average memory usage of the \textit{Single} scheme reaches 1035.04 MB, which may exceed the memory limits of most edge devices and could lead to premature termination of LM fine-tuning. RingAda also reduces memory usage compared to \textit{PipeAdapter}, as it avoids model staleness issues, meaning clients no longer need to store multiple versions of model parameters, as is required in traditional pipeline parallelism.

\section{Conclusion}
In this work, we have presented RingAda, a collaborative training framework compatible with edge devices for fine-tuning transformer-based large models using adaptable adapters. The core of RingAda lies in its ring topology for per-batch training, which has sequentially placed frozen transformer blocks and their associated trainable adapter modules on multiple edge devices, enabling fast and parameter-efficient fine-tuning. Through effective scheduling of layer unfreezing, RingAda deactivates the bottom-layer adapters and employs a novel pipeline-parallel training mechanism, utilizing early stopping of backward propagation at the lowest adapter layer to accelerate decentralized adapter fine-tuning. Our experiments have shown that RingAda can significantly reduce fine-tuning time and memory costs, demonstrating the potential for privacy-preserved and resource-efficient large model service provisioning in edge network environments.



\bibliographystyle{IEEETran}
\bibliography{Liang.bib}

\begin{thebibliography}{10}
\providecommand{\url}[1]{#1}
\csname url@samestyle\endcsname
\providecommand{\newblock}{\relax}
\providecommand{\bibinfo}[2]{#2}
\providecommand{\BIBentrySTDinterwordspacing}{\spaceskip=0pt\relax}
\providecommand{\BIBentryALTinterwordstretchfactor}{4}
\providecommand{\BIBentryALTinterwordspacing}{\spaceskip=\fontdimen2\font plus
\BIBentryALTinterwordstretchfactor\fontdimen3\font minus \fontdimen4\font\relax}
\providecommand{\BIBforeignlanguage}[2]{{%
\expandafter\ifx\csname l@#1\endcsname\relax
\typeout{** WARNING: IEEEtran.bst: No hyphenation pattern has been}%
\typeout{** loaded for the language `#1'. Using the pattern for}%
\typeout{** the default language instead.}%
\else
\language=\csname l@#1\endcsname
\fi
#2}}
\providecommand{\BIBdecl}{\relax}
\BIBdecl

\bibitem{touvron2023llama}
H.~Touvron, T.~Lavril, G.~Izacard, X.~Martinet, M.-A. Lachaux, T.~Lacroix, B.~Rozi{\`e}re, N.~Goyal, E.~Hambro, F.~Azhar \emph{et~al.}, ``Llama: Open and efficient foundation language models,'' \emph{arXiv preprint arXiv:2302.13971}, 2023.

\bibitem{brown2020language}
T.~Brown, B.~Mann, N.~Ryder, M.~Subbiah, J.~D. Kaplan, P.~Dhariwal, A.~Neelakantan, P.~Shyam, G.~Sastry, A.~Askell \emph{et~al.}, ``Language models are few-shot learners,'' \emph{arXiv preprint arXiv:2005.14165}, 2020.

\bibitem{Li2021infocom}
L.~{Li}, D.~{Shi}, R.~{Hou}, H.~{Li}, M.~{Pan}, and Z.~{Han}, ``To talk or to work: Flexible communication compression for energy efficient federated learning over heterogeneous mobile edge devices,'' in \emph{Proc. IEEE International Conference on Computer Communications (INFOCOM)}, Virtual Conference, May 2021.

\bibitem{li2023energy}
L.~Li, C.~Huang, D.~Shi, H.~Wang, X.~Zhou, M.~Shu, and M.~Pan, ``Energy and spectrum efficient federated learning via high-precision over-the-air computation,'' \emph{IEEE Transactions on Wireless Communications}, vol.~23, no.~2, pp. 1228--1242, 2023.

\bibitem{chen2022fedtune}
J.~Chen, W.~Xu, S.~Guo, J.~Wang, J.~Zhang, and H.~Wang, ``Fedtune: A deep dive into efficient federated fine-tuning with pre-trained transformers,'' \emph{arXiv preprint arXiv:2211.08025}, 2022.

\bibitem{shi2022toward}
D.~Shi, L.~Li, R.~Chen, P.~Prakash, M.~Pan, and Y.~Fang, ``Toward energy-efficient federated learning over 5g+ mobile devices,'' \emph{IEEE Wireless Communications}, vol.~29, no.~5, pp. 44--51, 2022.

\bibitem{xu2023FwdLLM}
M.~Xu, D.~Cai, Y.~Wu, X.~Li, and S.~Wang, ``Federated fine-tuning of billion-sized language models across mobile devices,'' \emph{arXiv preprint arXiv:2308.13894v2}, 2024.

\bibitem{devlin2018bert}
J.~Devlin, M.-W. Chang, K.~Lee, and K.~Toutanova, ``{BERT}: Pre-training of deep bidirectional transformers for language understanding,'' \emph{arXiv preprint arXiv:1810.04805}, 2018.

\bibitem{houlsby2019adapter}
N.~Houlsby, A.~Giurgiu, S.~Jastrzebski, B.~Morrone, Q.~De~Laroussilhe, A.~Gesmundo, M.~Attariyan, and S.~Gelly, ``Parameter-efficient transfer learning for {NLP},'' in \emph{Proc. of International Conference on Machine Learning (ICML)}, Long Beach, CA, June 2019.

\bibitem{cai2023adaFL}
D.~Cai, Y.~Wu, S.~Wang, F.~X. Lin, and M.~Xu, ``Efficient federated learning for modern {NLP},'' in \emph{Proc. of the Annual International Conference on Mobile Computing and Networking (MobiCom)}, Madrid, Spain, October 2023.

\bibitem{zhao2023fedprompt}
H.~Zhao, W.~Du, F.~Li, P.~Li, and G.~Liu, ``Fedprompt: Communication-efficient and privacy-preserving prompt tuning in federated learning,'' in \emph{Proc. of IEEE International Conference on Acoustics, Speech and Signal Processing (ICASSP)}, Rhodes Island, Greece, June 2023.

\bibitem{liao2024make}
B.~Liao, S.~Tan, and C.~Monz, ``Make pre-trained model reversible: From parameter to memory efficient fine-tuning,'' in \emph{Proc. of Advances in Neural Information Processing Systems (NIPS)}, Vancouver, Canada, December 2024.

\bibitem{ruckle2021adapterdrop}
A.~R{\"u}ckl{\'e}, G.~Geigle, M.~Glockner, T.~Beck, J.~Pfeiffer, N.~Reimers, and I.~Gurevych, ``Adapterdrop: On the efficiency of adapters in transformers,'' in \emph{Proc. of the Conference on Empirical Methods in Natural Language Processing (EMNLP)}, Punta Cana, Dominican Republic, November 2021.

\bibitem{liu2023improving}
C.~C. Liu, J.~Pfeiffer, I.~Vuli{\'c}, and I.~Gurevych, ``Improving generalization of adapter-based cross-lingual transfer with scheduled unfreezing,'' \emph{arXiv preprint arXiv:2301.05487}, 2023.

\bibitem{chen2024confidant}
Y.~Chen, Y.~Yan, Q.~Yang, Y.~Shu, S.~He, and J.~Chen, ``Confidant: Customizing transformer-based {LLM}s via collaborative edge training,'' \emph{arXiv preprint arXiv:2311.13381}, 2023.

\bibitem{vaswani2017attention}
A.~Vaswani, N.~Shazeer, N.~Parmar, J.~Uszkoreit, L.~Jones, A.~N. Gomez, {\L}.~Kaiser, and I.~Polosukhin, ``Attention is all you need,'' \emph{in Proc. of Advances in neural information processing systems (NIPS)}, December 2017.

\bibitem{narayanan2019pipedream}
D.~Narayanan, A.~Harlap, A.~Phanishayee, V.~Seshadri, N.~R. Devanur, G.~R. Ganger, P.~B. Gibbons, and M.~Zaharia, ``Pipe{D}ream: Generalized pipeline parallelism for {DNN} training,'' in \emph{Proc. of the ACM symposium on operating systems principles (SOSP)}, New York, NY, October 2019.

\bibitem{rajpurkar2016squad}
P.~Rajpurkar, ``Squad: 100,000+ questions for machine comprehension of text,'' \emph{arXiv preprint arXiv:1606.05250}, 2016.

\end{thebibliography}
\end{document}